\documentclass[pdflatex,sn-mathphys]{sn-jnl}

\usepackage{mathtools}
\usepackage[capitalize,nameinlink]{cleveref}
\usepackage{nicefrac}

\jyear{2023}%

\theoremstyle{thmstyleone}%
\newtheorem{theorem}{Theorem}
\newtheorem{proposition}[theorem]{Proposition}
\newtheorem{corollary}[theorem]{Corollary}%

\theoremstyle{thmstyletwo}%
\newtheorem{example}{Example}%
\newtheorem{remark}{Remark}%

\theoremstyle{thmstylethree}%
\newtheorem{definition}{Definition}%

\newcommand{\cvPt}[1]{\mathcal{#1}}

\DeclarePairedDelimiter{\abs}{\lvert}{\rvert}

\begin{document}

\title{Manifesting Unobtainable Secrets: Threshold Elliptic Curve Key Generation using Nested Shamir Secret Sharing}


\author*[1]{\fnm{Joanne L.} \sur{Hall}}
\email{joanne.hall@rmit.edu.au}

\author[2]{\fnm{Yuval} \sur{Hertzog}} 
\email{uv@tide.org}
\equalcont{These authors contributed equally to this work.}

\author[2]{\fnm{Michael} \sur{Loewy}}
\email{michael@tide.org}
\equalcont{These authors contributed equally to this work.}

\author[1]{\fnm{Matthew P.} \sur{Skerritt}}
\email{matt.skerritt@rmit.edu.au}
\equalcont{These authors contributed equally to this work.}

\author[2]{\fnm{Dominique} \sur{Valladolid}}
\email{dom@tide.org}
\equalcont{These authors contributed equally to this work.}

\author[1]{\fnm{Geetika} \sur{Verma}}
\email{geetika.verma@rmit.edu.au}
\equalcont{These authors contributed equally to this work.}

\affil*[1]{\orgdiv{School of Science}, \orgname{RMIT University}, \orgaddress{\street{LaTrobe St}, \city{Melbourne}, \postcode{3001}, \state{VIC}, \country{Australia}}}

\affil[2]{ \orgname{TIDE Foundation}, \orgaddress{\street{Level 2, 65-71 Belmore Road}, \city{Randwick}, \postcode{2031}, \state{NSW}, \country{Australia}}}


\abstract{We present a mechanism to manifest unobtainable secrets using a nested Shamir secret sharing scheme to create public/private key pairs for elliptic curves.  A threshold secret sharing scheme can be used as a decentralised trust mechanism with applications in identity validation, message decryption, and agreement empowerment. Decentralising trust means that there is no single point vulnerability which could enable compromise of a system.      Our primary interest is in twisted Edwards curves as used in EdDSA, and the related Diffie-Hellman key-exchange algorithms.  The key generation is also decentralised, so can be used as a decentralised secret RNG suitable for use in other algorithms.  The algorithms presented could be used to fill a ``[TBS]'' in the draft IETF specification ``Threshold modes in elliptic curves''  published in 2020 and updated in 2022}.




\maketitle

\section{Introduction} %

Digital signatures enable authority to be verified in a digital manner. A cryptographic signature is a digital signature in which encryption is used to ensure validity and nonrepudiation. Cryptographic digital signatures are used to verify identity \cite{ElectronicSignaturesAct2000}, signal approval \cite{ElectronicSignaturesAct2000}, and indicate authenticity \cite{schneider1996robust}. The modern world needs signature algorithms that are reliable and resilient to attack. 

The management of cryptographic keys is (and has always been) a significant challenge for any secure communications system \cite{menezes2021challenges}.  Key generation, distribution, storage, utilisation, and update are all long standing challenges for which new paradigms are required. This remains true even with the advent of quantum resilient cryptographic schemes.

Cryptographic algorithms use random number generation (RNG), usually for generating a secret. If the RNG has any bias or predictability, then any algorithm that uses that RNG may be vulnerable \cite{cohney2019too}. 

Threshold cryptography using our nested Shamir secret-sharing technique has robust RNG. When used for cryptographic signatures, there is no single-point  vulnerability that could enable total system compromise. Our work fills a gap in the current IETF draft specification for threshold modes \cite{draft-hallambaker-threshold-modes} (see \cref{sub:preliminaries:threshold}).

We use Shamir Secret Sharing \cite{shamir1979share} in a nested manner with a distributed system to develop an RNG scheme that is resistant to bias and accidental disclosure.  In particular, the distributed nature of the bias resistance mechanism makes it significantly more difficult for an attacker to bias the RNG.

Shamir secret-sharing \cite{shamir1979share} ensures that no party can act alone; multiple independent parties are required to participate to complete that algorithm.  Originally developed with the use case of sharing a decryption key, Shamir's secret sharing can  be used to create threshold cryptographic schemes that similarly require multiple independent parties. Threshold schemes can be thought of as analogous to a quorum of a committee. The threshold scheme described in this paper has been developed for the use case of user authentication, authorisation and authority management \cite{TIDE}, but could be applied in other circumstances.

Communication protocols, including cryptographic algorithms, are published as standards e.g. \cite{draft-hallambaker-threshold-modes,draft-hallambaker-threshold-sigs, RFC-7748, EdDSA-RFC, RandomNumbers-RFC}. 
The publication of technical standards means that vulnerabilities are likely to be found and, once found, will be exploited. It is not possible to remove all vulnerabilities; however the techniques we present in this paper allow us to ensure that there is no single point vulnerability which could enable system compromise. Consequently, a successful attack would need to simultaneously breach multiple independent systems.  

There are other distributed RNG and threshold key generation schemes which use nesting (though not called `nesting') described in recent literature \cite{cryptoeprint:2020/1390,Gennaro2007,kate2009distributed}.  Aumasson et.al.  in a survey of ECDSA threshold schemes \cite{cryptoeprint:2020/1390}, present a threshold key generation without analysis,  citation, or claim to originality.
Gennaro 2007 \cite{Gennaro2007} focusses on threshold generation of a secret which nominally can be used for key generation.  Their focus is on the uniformly random distribution of keys within the key space.    These distributed systems are dealerless, whereas our system, \cref{alg:distributed_eddsa_key_generation}, has a dealer. For more discussion see \cref{subsec:swarmdiscussion}.

Some draft standards are published whilst incomplete, allowing for the technical community to contribute to the development of standards. There are currently two IETF drafts relating to threshold schemes for elliptic curves, both by P. M. Hallam-Baker. The draft titled ``Threshold Modes in Elliptic Curves" \cite{draft-hallambaker-threshold-modes} (herein the \emph{``threshold modes IETF draft''}) is a broad description of threshold schemes for elliptic curves covering: Diffie-Hellman, Shamir secret sharing, and (briefly) signing.
It focusses on the cases of Curve25519/Ed25519 and Curve448/Ed448.
Conversely ``Threshold Signatures in Elliptic Curves'' \cite{draft-hallambaker-threshold-sigs} (herein the \emph{``threshold signatures IETF draft''}) deals specifically with a threshold signature scheme for Ed25519 and Ed448.

Another, more recent, draft, ``Two-Round Threshold Schnorr Signatures with FROST'' \cite{irtf-cfrg-frost-14}, covers threshold signatures, but not distributed key generation.

We note that the threshold signatures IETF draft has expired as of July 17\textsuperscript{th} 2021.   
The second-most-recent update of the threshold modes IETF draft (version 07 on April 22\textsuperscript{nd} 2022) was updated to remove some (but not all) references to the threshold signatures IETF draft.
Although the threshold modes IETF draft \cite[\textsection{5}]{draft-hallambaker-threshold-modes} does discuss EdDSA, it states that ``\dots threshold signatures are out of scope for this document \dots'' \cite[\textsection{}5.1]{draft-hallambaker-threshold-modes}.
As of the time of writing of this paper, none of the threshold signing algorithms from the threshold signatures IETF draft are present in the threshold modes IETF draft.

The key generation scheme we describe in this paper applies to both the threshold key exchange (using X25519 and X448), as well as the threshold digital signing (using EdDSA).
The primary interest of the authors is in the threshold signing; nonetheless, this paper is presented in full generality,  discussing both threshold signing and threshold key exchange.
Consequently, we consider the threshold signing algorithm described in Hallam-Baker's expired threshold signatures IETF draft to be relevant to this paper.

\subsection{Organisation of the paper}
We focus our attention on the elliptic curve based cryptographic schemes described in \cref{sub:elliptic_curve_cryptography}. A particular focus is on digital signatures, although the techniques we present (and the IETF drafts for which we fill a gap) also apply to threshold key exchange.

The paper is organised as follows: \cref{sec:technical_preliminaries} introduces technical preliminaries and definitions. \cref{sec:RNG} describes the nested Shamir secret sharing and swarm RNG. \cref{sec:EdDSA} contains EdDSA threshold scheme followed by \cref{sec:discussion} in which properties of nested Shamir secret sharing, swarm RNG and EdDSA threshold scheme are discussed. \cref{sec:conc} concludes the paper and presents some future directions.
 
\subsection{Authors Contribution}
Encryption alone does not address the problem of key distribution; Secret-sharing neither address who should run the sharing algorithm, nor how the shares should be distributed. In this paper, we discuss a novel nested secret sharing approach to digital signatures, where the dealer collects the shares from all of the parties who contribute their share to the secret.  This is a decentralised system with no single point of failure, even the dealer cannot know the secret without the contribution of a threshold number of parties.

The nested secret sharing scheme and digital signature algorithms described in this paper fill gaps in two IETF drafts  \cite{draft-hallambaker-threshold-modes, draft-hallambaker-threshold-sigs}.  The sections on {\em Verifiable Secret Sharing} \cite[\textsection{}4.3]{draft-hallambaker-threshold-modes} and {\em Distributed Key Generation} \cite[\textsection{}4.4]{draft-hallambaker-threshold-modes} state {``[TBS]''} (which we presume is shorthand for ``To be specified'').  Algorithm \ref{alg:distributed_eddsa_key_generation} could be used in the gap in the IETF draft \cite[\textsection{}4.4]{draft-hallambaker-threshold-modes}.

The nested secret sharing scheme described in Sections \ref{sec:RNG} and \ref{sec:EdDSA} of this paper, can be used as a decentralised unbiased RNG that is {\em fully secure}.  Formal entropy calculations are shown to demonstrate the unbiased RNG in the decentralised system.

\section{Technical Preliminaries}
\label{sec:technical_preliminaries}

\subsection{Security Models and Benchmarking}
  There are some common terminologies and concepts that are used to compare protocols and algorithms. 
This paper examines a multiparty protocol which can be used to generate a secret random number.

A useful way to measure the quality of an RNG is by measuring {\em entropy}.  We use Shannon entropy \cite{bierbrauer2016}, denoted $H(x)$, which is  different to thermodynamic entropy.
If an RNG 
has maximum entropy, then it is an {\em unbiased} RNG.


Using the associativity of addition, the addition of entropy can be generalised from $H(x+y)\geq \max\{H(x),H(y)\}$ \cite{madiman2008entropy}, to any number of discrete random variables:
Let $x_1, x_2, \dots x_n$ be discrete random variables on the same set which is closed under addition, then 
\begin{equation}\label{eqn:entropy_multiaddition}
   H(x_1+x_2+\dots+x_n)\geq \max \left\{H(x_1),H(x_2),\dots,H(x_n) \right\}.
\end{equation}

\Cref{eqn:entropy_multiaddition} is used in \cref{sec:discussion} to examine the proposed threshold key generation  algorithm.

An algorithm is {\em biased} if there is a systemic or repeatable way to influence the outcome of the algorithm. 
A multi-party computation is {\em unbiasable} if a corrupt party in the multiparty computation has a negligible probability of influencing the output of the computation.
The term {\em corrupt} may be used to describe a party which is dishonest, malicious, or faulty.    In many instances, the motivation of the corrupted party is not important, so a network failure is treated the same as maliciously withholding data.

A multi-party protocol is {\em fair} if an adversary can prematurely abort the computation, however, they do not gain any information by doing so.  
A multi-party computation is {\em fully secure} if no adversary can prevent the honest parties from obtaining their output.

Data security is often concerned with malicious changes to data, however we also need to guard against mundane adversaries such as hardware failure.  A multi-party computation has high {\em availability} if, with high probability, the multi-party computation can complete correctly.
Reconstruction of a secret via a secret sharing scheme is  a computation with high {\em availability}.
 
A general \emph{secret sharing scheme} divides a secret, \( \varsigma\), into \(n\) shares, which are distributed, one to each party such that  the shares do not reveal any information about the secret. In a $t$-in-$n$ secret sharing scheme,  the secret can be reconstructed if any $t$ or more of the parties combine  their shares together, but less than $t$ users can gain no information about the secret. The value $t$  is the {\em threshold}
of the secret sharing scheme.




\emph{Threshold cryptography} uses multi-party computation to complete a cryptographic algorithm.
The term ``threshold'' in this context is related to, but distinct from the threshold in secret sharing schemes.
In threshold cryptography it is possible to have an \(n\)-in-\(n\) threshold scheme.
The IETF drafts \cite{draft-hallambaker-threshold-modes,draft-hallambaker-threshold-sigs} refer to these cases as \emph{direct} threshold schemes.
Direct threshold does not require Shamir secret sharing, and so we are not concerned with the direct case for this paper.

\subsection{Shamir Secret Sharing}
\label{sec:inter}

The Shamir secret sharing scheme \cite{shamir1979share}  uses polynomials over $\mathbb{Z}_p$. 
There are several variations of the original Shamir secret sharing which provide for a variety of levels of robustness \cite{cheraghchi2019nearly,beimel2015protocols}, and features such as honest parties having the ability to detect corruption and abort \cite{cohen2022fairness}. 

In a $t$-in-$n$ Shamir secret sharing scheme a secret, \( \varsigma\), and a $t-1$ degree polynomial, \( P\in\mathbb{Z}_p[x] \), are generated such that \( P(0) = \varsigma \).
The  secret, \( \varsigma\), is shared in \(n\) parts, $\sigma_i := (x_i, y_i)$, $1<i<n$ which are points on the polynomial $P$.

Any \(t, (0 \le t \leq n)\) parties bring their shares together and use Lagrange interpolation to evaluate the polynomial.
That is, given \( C \subset \{1,\dotsc,n\} \) with \( \abs{C} \ge t \), the secret, \( \varsigma \), can be reconstructed by 
\[ \varsigma = \sum_{c \in C} y_c l_c(C) \]
where \( l_i(C) \) is the Lagrange coefficient for share, \( \sigma_i \), given the set of shares \( \{ \sigma_c \mid c \in C \}  \) calculated by
\[ l_i(C) = \prod_{c\in C, c \ne i} \frac{x_c}{x_c - x_i} .\]

We introduce nested Shamir Secret Sharing (\cref{sec:RNG}).

\subsection{Elliptic Curve Cryptography}
\label{sub:elliptic_curve_cryptography}

We are interested in the EdDSA cryptographic signature scheme \cite{Bernstein:2015,EdDSA-RFC,Bernstein:2012es}, and the related elliptic curve Diffie-Hellman key exchange scheme \cite{Bernstein:2006kw,EdDSA-RFC}. The unifying aspect of these is twisted Edwards curves which we introduce in \cref{ssub:preliminaries:twisted_edwards_curves}.

\subsubsection{Conventions and Notation}

The cryptographic schemes we consider use an elliptic curve, \( E(x,y) \),  over a finite field, \( \mathbb{F} \). The points of $E(x,y)$ form a group under point addition, see e.g. 
\cite{Bernstein:2015,EdDSA-RFC,Bernstein:2012es,Hamburg:2015,Bernstein:2008,Edwards:2007,Bernstein:2006kw}. The schemes operate within a cyclic subgroup, of prime order with a known generator. 

We use the following notational conventions:
    \begin{itemize}
        \item Elliptic curve points are denoted in a calligraphic font (e.g., \( \cvPt{P} \)).
        \item Elements of the finite field \( \mathbb{F} \) are referred to as \emph{scalars}.
        \item The \emph{``base point''}, denoted by \( \cvPt{G} \),  is the generator of the prime order subgroup. 
        \item The order of the prime order subgroup is denoted by \( \ell \).
        \item Repeated addition of a curve point is denoted multiplicatively. For example \( 2\cvPt{P} = \cvPt{P} + \cvPt{P} \), and \( s\cvPt{P} = \cvPt{P} + \dotsb + \cvPt{P} \).
    \end{itemize}
    

\subsubsection{Twisted Edwards Curves}
\label{ssub:preliminaries:twisted_edwards_curves}

Twisted Edwards curves were introduced by Bernstein et al. \cite{Bernstein:2008} in 2008 as a generalisation of Edwards Curves \cite{Edwards:2007}. 
Twisted Edwards curves are useful as they have a group addition law which is fast and works on all pairs of inputs (unlike other curves which can require special handling of the identity element).
In addition, every elliptic curve that has a Montgomery form \cite{Okeya_et_al:2000} can be shown \cite{Bernstein:2012es} to be bi-rationally equivalent to some twisted Edwards curve, so existing schemes can  leverage the desirable properties of twisted Edwards curves.

The EdDSA scheme is general in the sense that it is suitable for use with any twisted Edwards curve, so long as the required parameter constraints are satisfied \cite{Bernstein:2015}. EdDSA is realised in the commonly used Ed25519 \cite{Bernstein:2012es} and the less commonly used Ed448 \cite{Hamburg:2015}.  Details suitable for practical implementation of these two variants can be found in IEEE RFC8032 \cite{EdDSA-RFC}.


The elliptic curve used for Ed25519 is bi-rationally equivalent to Curve25519, introduced for elliptic curve Diffie-Hellman key exchange in 2006 \cite{Bernstein:2006kw}.  The theory presented in the Curve25519 paper \cite[Thm 2.1]{Bernstein:2006kw} is more general than just Curve25519,  applying to other curves of a similar form. Indeed, RFC-7748 \cite{RFC-7748} allows for Diffie-Hellman key exchange using both the Ed25519 curve and the Ed448 curve. 

\begin{definition}[Curve25519, X25519, Curve448, X448]\label{defn:Curve_and_X}
    Let \emph{Curve25519} denote both the twisted Edwards curve used for Ed25519 and the bi-rationally equivalent Montgomery form of that curve. Let  \emph{X25519} denote the function for performing scalar point multiplication on the Montgomery form of Curve25519.
    
    Similarly, let \emph{Curve448} denote both the Edwards curve used for Ed448 and the bi-rationally equivalent Montgomery form of that curve. Let \emph{X448} denote the function for performing scalar point multiplication on the Montgomery form of Curve448.
\end{definition}

The terminology in \cref{defn:Curve_and_X} aligns with RFC-7748 \cite{RFC-7748}. The reader should be aware that Bernstein's original paper \cite{Bernstein:2006kw}  used the name ``Curve25519'' to refer to what we (and the RFC) call ``X25519''.

\begin{definition}[secret scalar, $ a $] \label{def:secret_scalar}
    Both X25519/X448 and EdDSA have a secret value that is used to scale the base point to produce the elliptic curve point that creates the public key. We call this secret value the \emph{secret scalar}, and denote it by \( a \).
\end{definition}

The EdDSA scheme creates private keys by generating a random secret (named \(k\) in the RFC \cite{EdDSA-RFC}) which is hashed, and then split in half; the first half becomes the private key (after Key Clamping; see \cref{ssub:preliminaries:keyclamping}) the second half is used in signing. 
During signing, the second half of the hash of $k$ is appended to the message and hashed to produce a value (named \(r\) in the RFC \cite{EdDSA-RFC}).

In this document we use the following terminology for the EdDSA specific values.
\begin{definition}[secret seed, secret prefix, secret signing nonce] \label{def:secret seed}\label{def:secret prefix}\label{def:secret signing nonce}
\;
    \begin{itemize}
        \item The original randomly generated value, \(k\), is the \emph{secret seed}.
        \item The first half of the hash of the secret seed is the \emph{secret scalar} (consistent with \cref{def:secret_scalar}).
        \item The second half of the hash of the secret seed is the \emph{secret prefix}.
        \item The value, \( r \), used in signing is the \emph{secret signing nonce}\footnote{A nonce is a value which is used exactly once, then discarded.}.
    \end{itemize}
\end{definition}

\subsubsection{Key Clamping}
\label{ssub:preliminaries:keyclamping}

Private keys for both the signing and key exchange algorithms are byte strings, interpreted as scalar values from a finite field, \( \mathbb{F}_p \), where \( p = 2^{255}-19 \) for Curve25519, and \( p = 2^{448} - 2^{224} - 1 \) for Curve448.

Both Curve25519 and Curve448 use {\em key clamping}\footnote{The term `key clamping' is widely used in practice, though not in the academic literature.}, see for example \cite{Bernstein:2006kw,Bernstein:2012es,Bernstein:2015,RFC-7748,EdDSA-RFC}. 
The key clamping procedure 
sets specific bits to 0 or 1 during key creation (after randomly generating the bit-sting for use as the private key).

In the case of Curve25519, the key clamping has the effect of guaranteeing that a private key is in the set 
\[ \mathit{KeySpace}_{25519} := \left\{ n \in \mathbb{F}_p \mid n = 2^{254} + 8k \text{ for } 0 \le k \le 2^{251}-1 \right\}, \]
and Curve448 key clamping ensures the private key is in the set 
\[ \mathit{KeySpace}_{448} := \left\{ n \in \mathbb{F}_p \mid n = 2^{447} + 4k \text{ for } 0 \le k \le 2^{445}-1 \right\}. \]
In particular, Curve25519 private keys are multiples of 8 and Curve448 private keys are multiples of 4.

Bernstein \cite{Bernstein:2006kw} explains that the multiple of 8 for a Curve25519 key acts as a protection against a small subgroup attack of the Diffie-Hellman key exchange. Such an attack would use a low order Curve25519 point 
to discover three bits of the private key.

EdDSA keys also use clamping; however, 
there does not appear to be an analogous small subgroup attack in the EdDSA signing algorithm.
Galbraith in his 2020 CRYPTREC external evaluation report of EdDSA \cite[\textsection 6.1]{Galbraith:2020} opines that he does not know why the EdDSA specification uses key clamping.
He notes that key clamping results in a non-uniform sampling of the set \( \mathbb{Z}_\ell \), and complicates security proofs, but does not appear to weaken security. We note that key clamping has the additional effect of giving a one-to-one correspondence between unreduced clamped keys and keys reduced modulo \( \ell \).

It appears that the inclusion of Key Clamping in Ed25519 is motivated by the choice of curve in the original Ed25519 paper by Bernstein et al. \cite{Bernstein:2008}.  The curve chosen for Ed25519 is bi-rationally equivalent to Curve25519 (see \cref{ssub:preliminaries:twisted_edwards_curves}) introduced in  Bernstein's earlier paper \cite{Bernstein:2006kw}. Observe that the inclusion of key clamping in the Ed25519 scheme ensures that all Ed25519 keys produced according to the specification \cite{Bernstein:2008, EdDSA-RFC}
are bi-rationally equivalent to safe keys for X25519 and vice versa. This similarly holds for X448 and Ed448.
In other words it allows signing keys to safely also be used for key exchange, although we note that doing so would violate  the key separation principle of not using the same key for multiple purposes \cite{gligoroski2008importance}.



\subsection{Threshold Key Exchange and Threshold Signing }
\label{sub:preliminaries:threshold}

Threshold cryptographic techniques are multi-party computations performed in parallel, such that the cryptographic result is a combination of the contributions from a threshold number of parties. The combined result is referred to as the \emph{``aggregate''} result (e.g., aggregate public key, aggregate signature).


The threshold modes used in this paper require two types of parties: one dealer and some actors.
The dealer requests the key-exchange or signature calculation, and the actors perform the multi-party threshold calculation.
In the case of threshold digital signatures, the actors are usually called signers.
Conceptually the dealer is separate to the actors but   the dealer could  also participate as one of the actors with security and logistical implications dependent on context.

In this paper we present a threshold key generation scheme using nested Shamir secret sharing.
Our scheme is usable with the threshold singing \cite{draft-hallambaker-threshold-sigs}, and the threshold key exchange \cite{draft-hallambaker-threshold-modes} approaches described by Hallam-Baker.

\subsubsection{Key Exchange}
The IETF draft specifications \cite{draft-hallambaker-threshold-modes, draft-hallambaker-threshold-sigs}  do not give an explicit algorithm for threshold key exchange. 
Generic key combination and result combination laws are explained \cite[\textsection{3}]{draft-hallambaker-threshold-modes} 
with a statement that the relevant operators for elliptic curve cryptosystems are point addition \cite[\textsection{5}]{draft-hallambaker-threshold-modes}.
However, the specifics of applying threshold key exchange to elliptic curves is only implied.
Similarly, Shamir secret sharing is explained in general in the threshold modes IETF draft \cite[\textsection{4}]{draft-hallambaker-threshold-modes}, but it's application to  elliptic curve Diffie-Hellman is only implied.

We outline the mathematical particulars of the Diffie-Hellman exchange:
Let $\mathbb{F}$ be a finite field.  Suppose we have a secret scalar, \( a\in\mathbb{F} \), that is shared via a \(t\)-in-\(n\) Shamir secret sharing scheme with shares \( \sigma_1, \dotsc, \sigma_n \) (where each \(\sigma_i = (x_i, y_i)\in\mathbb{F}\times\mathbb{F} \)). 
Suppose that we wish to establish a shared secret, \( \cvPt{K}\in\mathbb{F} \), with an entity whose public key is \( \cvPt{P}\in\mathbb{F} \).
The general Diffie-Hellman calculation of the shared secret
  $  \cvPt{K} = a \cvPt{P}$
can be performed by a set of  shares \( \left\{  \sigma_c \mid c \in C \subset \{1, \dotsc, n \}, \vert C\vert \geq t \right\} \) by
\begin{equation*}
    \cvPt{K} = \sum_{c \in C} l_c(C) \cvPt{K}_c
\end{equation*}
where $\cvPt{K}_c = y_c \cvPt{P}$ and $l_c(C)$ is the Lagrange coefficient.  

Note that if the individual actors know their Lagrange coefficients, then they can calculate \( l_c(C)\cvPt{K}_c \) directly, and the dealer then sums the results together.
Alternatively, the actors can calculate \( \cvPt{K}_c \), then the dealer can apply the Lagrange coefficients and sum the results.
We discuss  these options  in \cref{sub:threshold_keygen:encryption_keys_and_actor_identity}.

\subsubsection{Digital Signatures} \label{ssub:preliminaries:threshold:digital signatures}
The threshold EdDSA algorithms described in the threshold signatures IETF draft \cite{draft-hallambaker-threshold-sigs} differ slightly from the direct EdDSA algorithms in the RFC \cite{EdDSA-RFC}.
These differences are due to the multi-party nature of threshold cryptography. 
We outline these here; the reader should consult the relevant RFCs for more detailed instructions.

Threshold Signing using the threshold signatures IETF draft \cite{draft-hallambaker-threshold-modes} protocol is a two-part process:
\begin{enumerate}
    \item Signer $i$'s contribution  is calculated as \( \cvPt{R}_i= r_i \cvPt{G} \) (using their own secret signing nonce \( r_i \)).
    These contributions are sent  to the dealer.
    \item Each signer receives \( \cvPt{R} = \sum_j \cvPt{R}_j \) and thier own Lagrange coefficient \( l_i(C) \) from the dealer.\footnote{It's possible that not all signers will respond to the initial request, so these coefficients must be calculated after receiving responses.}
    The signers then complete the signature as per  the RFC \cite{EdDSA-RFC}, with the one difference being that signer $i$'s contribution of \( S \) is calculated as \( S_i = r_i + l_i(C) k a_i \).\footnote{This differs from the usual signature calculation of \( S = r + ka \) by extra inclusion of the Lagrange coefficient \( l_i(C) \) as a coefficient of the secret scalar.}
    These contributions are sent back to the dealer.
\end{enumerate}
Upon receiving contributions \( S_i \) from all signers, the dealer calculates the aggregate value \( S  = \sum_j S_j \), tests that the signature \( (\cvPt{R},S) \) is valid, and if valid, publishes the signature.


The deterministic method of calculating the secret signing nonce 
\cite[\textsection{}5.1.5]{EdDSA-RFC} is insecure when used for threshold signing \cite{draft-hallambaker-threshold-sigs}; the dealer may exploit it to discover  secret scalars by requesting signatures of the same message using different values of \( \cvPt{R} \) or of \( \sigma_i \)
\cite[\textsection{3.4.1}]{draft-hallambaker-threshold-sigs}.

The insecurity of a deterministic nonce \cite{draft-hallambaker-threshold-sigs}  is due to the dealer sending both the aggregate value of \( \cvPt{R} \) as well as the Lagrange coefficient \( l_i(C) \) to each signer.
As such, the dealer controls  the coefficient \( \mu = l_i(C) k \) of the secret scalar \( s_i \) used in the signature calculation by the signer.\footnote{Noting that the value \( k \) is a hash of the values \( \cvPt{R} \), \( \cvPt{A} \), and the message \( M \), all of which are known to the dealer.}
If the signers use a deterministic secret signing nonce \( r_i \), and the dealer requests two signatures of the same message from a signer, using either a different value of \( \cvPt{R} \), or a different Lagrange coefficient \( \sigma_i \), then the dealer need only solve the  simultaneous linear equations
\begin{align*}
    S_1 &= r_i + \mu_1 s_i \mod \ell \\
    S_2 &= r_i + \mu_2 s_i \mod \ell
\end{align*}
in order to recover both the secret signing nonce and (more importantly) the secret scalar.

As a consequence of the insecurity of a deterministic nonce, Hallam-Baker \cite{draft-hallambaker-threshold-sigs} requires that secret signing nonces are randomly generated.
Consequently, the threshold signatures are not deterministic (unlike the usual EdDSA \cite{EdDSA-RFC} signatures which are); the same message signed twice by the same signers have (almost certainly) different valid signatures.

The threshold signatures IETF draft claims \cite[\textsection{3.4.1}]{draft-hallambaker-threshold-sigs} (but provides no references) that all known approaches to deterministic nonce generation are 
vulnerable to rogue key and malicious contributions attacks.

There is some discrepancy between the two IETF drafts in regards to secret prefixes and the calculation of \( r_i \).
The threshold modes IETF draft \cite{draft-hallambaker-threshold-modes}  suggests---specifically for Ed25519 and Ed448---ways in which an aggregate secret prefix be calculated \cite[\textsection{5.1.1--5.1.2}]{draft-hallambaker-threshold-modes}.
However, the requirement of random secret signing nonces \cite{draft-hallambaker-threshold-sigs} removes the need for a secret signing prefix.

\subsubsection{Attacks on Threshold Elliptic Curve Cryptography}
\label{ssub:Preliminary:ECC:Threshold Sigs and KEx:Attacks}
Two attacks on a threshold system are described in the IETF drafts: the \emph{rogue key attack} \cite[\textsection{}3.4.4]{draft-hallambaker-threshold-sigs} \cite[\textsection{}7.2]{draft-hallambaker-threshold-modes}, and the \emph{malicious contribution attack} \cite[\textsection{}3.4.3]{draft-hallambaker-threshold-sigs}. The malicious contribution attack in the IETF drafts is not of security concern as it merely causes the threshold calculation to fail \cite{draft-hallambaker-threshold-sigs}, necessitating a rerun. 
Analogous malicious contributions during key generation can cause key generation to fail, but do not compromise the security of the generated key; we discuss these in \cref{ssub:discussion_keygen:malicious contribution}.
The rogue key attack, however, is directly relevant to the security of the scheme we present in this paper.

\paragraph{Rogue Key Attack}
The rogue key attack is possible if an actor can hold back its contribution to the key generation until it has seen the contributions from all other actors.
In this case the corrupt actor can force the value of the aggregate public key to be any curve point they wish.
In particular, they can force it to be a public key for which only they know the secret key \cite[\textsection{}3.4.4]{draft-hallambaker-threshold-sigs}.

If a rogue key attack is successful, then the actor that performed it (and only that actor) will know the secret key corresponding to the aggregate public key.
Moreover, the aggregate secret key will not match the aggregate public key.
Correspondingly,  threshold operations will complete without error, but they will not produce the correct result.

We note that a threshold cryptographic failure stemming from a rogue key attack might not be immediately obvious.
Indeed, a failed signature will be immediately obvious (so long as the signer checks the signature before publishing it), but a threshold key exchange for message encryption\footnote{As opposed to message decryption for which a failure will likely be immediately obvious.} might not be immediately noticed.

The threshold modes IETF draft \cite{draft-hallambaker-threshold-modes} introduces the rogue key attack in the context of actors bringing their own (presumably pre-generated) keys together in an ad-hoc manner to create an aggregate key \cite[\textsection{}3.4.4]{draft-hallambaker-threshold-sigs}.
It is claimed that:
\begin{quote}
    ``Enabling the use of threshold signature techniques by ad-hoc groups of signers using their existing signature keys as signature key shares presents serious technical challenges that are outside the scope of this specification.'' \cite[\textsection{3.4.4}]{draft-hallambaker-threshold-sigs}
\end{quote}
We note,  that the scope presented by this quote is unnecessarily narrow.
Keys need not be pre-existing, and the groups need not be ad-hoc.
Any system in which actors provide their own key contributions (whether generated for the purpose like our system, or pre-existing) is potentially vulnerable to a rogue key attack.

The rogue key attack is fundamentally a key generation attack.
The IETF signatures draft states that the rogue key attack is not relevant to the threshold signature scheme presented therein \cite[\textsection{}3.4.4]{draft-hallambaker-threshold-sigs};
we interpret this to be because one cannot launch a rogue key attack against a signature algorithm.

For a rogue key attack to be exploitable, the aggregate public key must be published or otherwise distributed in an identical manner to any other public key for the system in question.
If a key is thus published, then it is likely (but not necessarily) the case that the rogue key attack was not detected.
We discuss the particulars around this in \cref{ssub:publicatoin_and_use_of_public_key}.

If multiple corrupt actors attempt a rogue key attack they will either all fail (because each malicious actor must wait to receive the contributions from all other actors) or only one will succeed (the actor who waits the longest).

Our key generation algorithm (presented in \cref{sec:EdDSA}) is not susceptible to the rogue key attack.

\section{Nested Shamir Secret Sharing}\label{sec:RNG}
We introduce a vehicle for random number generation (RNG), whereby the random number generated remains perpetually hidden. The suggested mechanism is nested Shamir secret sharing, a variation of Shamir secret sharing   \cite{shamir1979share}.  Secret random numbers are important as keys and nonces in cryptographic algorithms. 

\subsection{Swarm RNG}

Distributed random number generators use multi-party computation to generate random numbers. There are multi-party coin flipping protocols which generate a stream of random bits \cite{cleve1986limits,beimel2015protocols}. 

The Swarm RNG we propose uses nested Shamir secret sharing to generate a random number within a given range.

We explore nested secret sharing as an efficient method for bias-resistant RNG for use in signing algorithms (further details in \cref{sec:EdDSA}).
Applications (discussed in \cref{sec:discussion}) in Distributed Key Generation such as threshold signing and threshold key exchange use variations on  Shamir secret sharing  \cite{shamir1979share}.

We assume a system consisting of a large number of potential parties. Let a subset $P=\{p_1,\dots, p_n\}$ of parties, be called a {\em{swarm}}, and let $d$ denote the dealer.
   The swarm together with the dealer generates a secret, \( \varsigma\in\mathbb{Z}_q\) (for some prime \(q\)), and a \(t\)-in-\(n\) Shamir Secret Sharing polynomial, \( P \in\mathbb{Z}_q[X]\), such that \( P(0) = \varsigma \) in such a way that no individual member of the swarm ever sees \( P \) nor \( \varsigma \).
 
   Each party, \( p_i \) in the swarm precisely possesses:
    \begin{enumerate}
        \item A secret \( 2 \le \varsigma_i < q \) which it has randomly generated. 
        \item A secret degree-(\(t-1\)) polynomial, \( P_i \), which it has  randomly generated so that \( P_i(0)  = \varsigma_i \) (i.e., a Shamir Secret Sharing polynomial for \( \varsigma_i \)).
        \item A known \( x \)-coordinate \(x_i\) (known to all parties in the system.)
        \item The set $X=\{x_1, \dots, x_n\}$ of $x$ values for each member of the swarm.
    \end{enumerate}
    From these each swarm party $p_i, 1<i<n$  generates shares \( (x_j, P_i(x_j)) \) for each other party, \( p_j, j\neq i \) in the swarm. These shares are encrypted so that only \( p_j \) can decrypt the share \( (x_j, P_i(x_j)) \) and are then  sent to the dealer. After the dealer has received an encrypted share from \emph{all} the swarm parties, they then send each encrypted share to its intended swarm party.
    
    \begin{definition}[$\sigma_{i \to j}$]
        We denote by 
        \[ \sigma_{i \to j} := (x_j, P_i(x_j)) \]
        the share of secret \( \varsigma_i \) (generated by participant \( p_i \)) that is sent to participant \( p_j \).
    \end{definition}
    
    Each party, \( p_i \), in addition to items $1-4$, collects  \( n \) shares \( \sigma_{k \to i} = (x_i, P_k(x_i)) \) for \( 1 \le k \le n \). Note that this includes a share from its own polynomial.
    
    From this information each party calculates
    \begin{equation}
        y_i := \sum_{k=1}^n P_k(x_i) \mod p
    \end{equation} 
    and thus party $p_i$ has its share, $\sigma_i := (x_i, y_i)$, of the polynomial $  P = \sum_{k=1}^n P_k(x_i)$.
    Consequently,
    \begin{equation}
     \varsigma = P(0) = \sum_{k=1}^n P_k(0) . 
     \end{equation}
     Note that  no party (neither the dealer, nor the swarm parties)  ever see the secret \( \varsigma\) nor the polynomial \( P \).

    Once the swarm parties have generated and distributed the shares obtained by using Equations \(1\), \(2\) and \(3\),  Lagrange interpolation of the polynomial, \( P \), can reconstruct the secret, $\varsigma$, using  any \( t \) or more of the shares \( \sigma_k \).
    
    That is, given \( C \subset \{1,\dotsc,n\} \) with \( \abs{C} \ge t \) the secret, \( \varsigma \), can be reconstructed by 
    \[ \varsigma = \sum_{c \in C} y_c l_c(C) \]
    where \( l_i(C) \) is the Lagrange coefficient for party \( p_i \). 

\section{Distributed Threshold Key Generation}
\label{sec:EdDSA}

We leverage the nested Shamir secret sharing scheme (\cref{sec:RNG}) to fill the ``[TBS]'' for Distributed Key Generation in Hallam-Baker's IETF draft \cite[\textsection{}4.4]{draft-hallambaker-threshold-sigs}.Our key generation algorithm is described in \cref{alg:distributed_eddsa_key_generation} on \cpageref{alg:distributed_eddsa_key_generation}.

\begin{algorithm}
    \caption{Distributed Key Generation}
    \label{alg:distributed_eddsa_key_generation}
    \begin{algorithmic}[1]
        \Require Twisted Edwards curve subgroup with generator \( \cvPt{G} \) of prime order \( \ell \)
        \Require Values \(b\), \(c\), and hash function \( H \) (as per EdDSA \cite{EdDSA-RFC,Bernstein:2015}).
        \Require Functions \( \operatorname{encrypt}_1, \dotsc, \operatorname{encrypt}_n \) and  \( \operatorname{decrypt}_1, \dotsc, \operatorname{decrypt}_n \)
        \Require \( \tau \) maximum wait time in seconds
        \Repeat
            \State Dealer selects \( n \) actors, \( \{s_1, \dotsc, s_{n}\} \subset S \).
            \State Dealer ascertains \( x_1, \dotsc, x_n \) for actors \( s_i, \dotsc, s_n \) \Comment{See \cref{sub:EdDSA:x_coordinates}}
            \State Dealer sends \( (x_1, \dotsc, x_n) \) to all actors.\label{alg:finalising_key_generation:send_x_coordinates} \label{alg:distributed_eddsa_key_generation:send_x_coords}
            \State Each actor \( s_i \) performs \Call{Begin}{$x_1, \dotsc, x_n$} \Comment{In parallel}
            \State Dealer waits for \( \tau \) seconds.  
            \State Let $c$ denote the number of complete responses received from actors
        \Until{ \( c = n \) }\label{alg:distributed_eddsa_key_generation:wait for actors}
        \State Dealer sends \( (\mathfrak{M}_{1 \to i}, \dotsc, \mathfrak{M}_{n \to i}) \) to each actor \( s_i \)\label{alg:distributed_eddsa_key_generation:distribute encrypted messages}
        \State Each actor \( s_i \) performs \Call{Complete}{$ \mathfrak{M}_{1 \to i}, \dotsc, \mathfrak{M}_{n \to i}$  } \Comment{In parallel}

        \Statex\vspace{-0.75\baselineskip}
        \Procedure{Begin}{$x_1, \dotsc, x_n$}\label{proc:begin}
            \State \( (a_i, p_i) \gets \Call{NewSecretScalar}{ } \)
            \State \( r_i \gets H(p_i) \)
            \State \( \left(\cvPt{A}_i,\cvPt{R}_i\right) \gets \left(a_i\cvPt{G},r_i\cvPt{G}\right) \)
            \ForAll{ \( k \in \{ 1, \dotsc, t-1 \} \) }
                \State Randomly generate \( 0 \le \lambda_k < \ell \)
            \EndFor
            \State Define \( P_i(x) := (\lambda_{t-1} x^{t-1} + \dotsc + \lambda_1 x + a_i) \!\mod \ell \) 
            \ForAll{ \( j \in \{ 1, \dotsc, n \} \) }
                \State \( \sigma_{i \to j} \gets P_i(x_j) \)\label{alg:preliminary_key_generation:calculate_share}
                \State \( S_i \gets \left(H\!\left(\cvPt{A}_i, \cvPt{R}_i\right)\,a_i + r_i\right) \mod \ell \)  \label{alg:preliminary_key_generation:create_zk_proof}
                \State \( \mathfrak{M}_{i \to j} \gets \operatorname{encrypt}_j(\sigma_{i \to j}, \cvPt{R}_i, \cvPt{A}_i, S_i) \) 
                \label{alg:preliminary_key_generation:encrypt_share}
            \EndFor
            \State Send \( (\mathfrak{M}_{i \to 1}, \dotsc, \mathfrak{M}_{i \to n}) \) to the dealer.
        \EndProcedure

        \Statex\vspace{-0.75\baselineskip}
        \Procedure{Complete}{$ \mathfrak{M}_{1 \to i}, \dotsc, \mathfrak{M}_{n \to i}$}
            \ForAll{ \( j \in \{ 1, \dotsc, n \} \) }
                \State \( \left(\sigma_{j \to i}, \cvPt{R}_j, \cvPt{A}_j, S_j\right) \gets \operatorname{decrypt}_i(\mathfrak{M}_{j \to i}) \) \label{alg:finalising_key_generation:decrypt_share}\label{alg:finalising_key_generation:check S}
                \State Check: \( 2^c\cvPt{A} \ne \cvPt{O} \) \label{alg:finalising_key_generation:checks:low order curve point}
                \Comment{Abort if check fails}
                \State Check: \( 0 \le S_j < \ell \)  \label{alg:finalising_key_generation:checks:S}
                \Comment{Abort if check fails}
                \State Check: \( S_j\cvPt{G} = H\!\left(\cvPt{A}_i, \cvPt{R}_i\right)\cvPt{A}_j + \cvPt{R}_j\) \label{alg:finalising_key_generation:checks:verification equation}
                \Comment{Abort if check fails}
            \EndFor
            \State \( \sigma_i \gets \sum_{j=1}^n \sigma_{j \to i} \)

            \State \( \cvPt{A} \gets \sum_{j=1}^n  \cvPt{A}_{j} \)

            \State Store \( \cvPt{A} \) and \( \sigma_i \)
            \State Send \( \cvPt{A} \) to the dealer.
        \EndProcedure
        
        \Statex\vspace{-0.75\baselineskip}
        \Procedure{NewSecretScalar}{ }
        \Comment{See discussion in \cref{sub:EdDSA:random_generation_of_secret_scalar}}
        \algnotext{EndProcedure}
        \EndProcedure
    \end{algorithmic}
\end{algorithm}

At the end of  \cref{alg:distributed_eddsa_key_generation}, the secret key is the aggregate secret scalar, \( a = \sum_{i=1}^n a_i \). Each actor, \( s_i \), has a Shamir secret share, \( \sigma_i \), of the key, $a$, but no party has ever seen the aggregate key, $a$.

With a key generated by \cref{alg:distributed_eddsa_key_generation}, signing and key exchange may be performed  as described in the IETF drafts \cite{draft-hallambaker-threshold-sigs,draft-hallambaker-threshold-modes}.

Note that encrypting the share, \( \sigma_{i \to j}\), prevents the dealer from learning the secret, $a$.
Actors never see the private key contributions from the other actors, they see only their own private key contribution, the public key contributions from the other actors, and the final aggregate public key.

Key generation via nested Shamir secret sharing avoids any party directly generating or requesting the key. 
With such a direct request, the party making the request  knows the secret, even though the swarm (ideally) would not, which would be a single point of failure.
Instead, we directly incorporate the swarm RNG into the key generation.
The nested Shamir secret sharing is thus an integral part of the key generation.

The rogue key attack \cite[\textsection{}3.4.4]{draft-hallambaker-threshold-sigs} (discussed in \cref{sub:preliminaries:threshold}) is avoided in two steps.
Firstly, the dealer must wait (\cref{alg:distributed_eddsa_key_generation} \cref{alg:distributed_eddsa_key_generation:wait for actors}) to receive all contributions from the actors before distributing the encrypted messages.
This alone prevents the rogue key attack if the dealer is honest.
A dishonest dealer, however, may nonetheless perform a rogue key attack by also participating as one of the actors, or by colluding with a corrupt actor. Note that waiting is necessary to ensure that the dealer receives responses from the full swarm.

To protect against a malicious dealer we ensure that each actor, \( s_i \), provides a zero-knowledge proof of their knowledge of the secret scalar, \( a_i \), corresponding to their public key contribution, \( \cvPt{A}_i \).
The proof takes the form of an EdDSA signature of an empty message (\cref{alg:distributed_eddsa_key_generation}~\cref{alg:preliminary_key_generation:create_zk_proof}) with corresponding verification (\cref{alg:distributed_eddsa_key_generation}~\cref{alg:finalising_key_generation:checks:low order curve point,alg:finalising_key_generation:checks:S,alg:finalising_key_generation:checks:verification equation}).
We discuss the rogue key protections and the zero-knowledge proof in \cref{ssub:discussion_keygen:rogue key attack prevention}.

We note that \cref{alg:distributed_eddsa_key_generation}, as presented, has the dealer select \( n \) new actors in the event that some actors do not respond in time.
The new set of actors may include actors previously selected (although it is prudent not to re-select unresponsive actors).
For the generality of the algorithm,  the \( x \) coordinates are re-sent to the full set of actors on  \cref{alg:finalising_key_generation:send_x_coordinates}.
In practice, it might be possible to only send the list of \( x \) coordinates to the subset of new actors.
In short, if the coordinates are tied to the actors then all actors will need to re-perform the \textsc{Begin} procedure (\cref{alg:distributed_eddsa_key_generation}, \cref{proc:begin}); see discussion 
in \cref{sub:EdDSA:x_coordinates}.

\subsection{Random Generation of the Secret Scalar}
\label{sub:EdDSA:random_generation_of_secret_scalar}
 The \Call{NewSecretScalar}{} function (\cref{alg:generate_secret_scalar}) that is performed by the actors to randomly generate a secret scalar does not specify the random generation method.
As discussed under ``Digital Signatures'' in \cref{sub:preliminaries:threshold}, using a deterministic nonce is insecure, and so the secret prefix (see \cref{def:secret prefix}) is unnecessary.


For the sake of consistency with existing literature and implementations, it  seems best to follow the existing specifications \cite{Bernstein:2006kw,Bernstein:2012es,Bernstein:2015,EdDSA-RFC} as closely as possible.
Furthermore, the astute reader will have observed that the \Call{Begin}{} procedure in \cref{alg:distributed_eddsa_key_generation} expects to receive two values from \Call{NewSecretScalar}{}, the 2\textsuperscript{nd} of which is used exactly once for the zero knowledge proof.
Since we do not need the secret prefix for signing (and since key exchange never had a secret prefix), we propose to follow the EdDSA specification for key generation, but to re-purpose the secret prefix for the zero knowledge proof.
Consequently, we present \cref{alg:generate_secret_scalar} as the best implementation of \Call{NewSecretScalar}{}.

\begin{algorithm}
    \caption{\(\operatorname{NewSecretScalar} \)}
    \label{alg:generate_secret_scalar}
    \begin{algorithmic}[1]
        \Require Cryptographically secure random number generator.
        \Require Values \(b\), \(c\), and hash function \( H \) (as per EdDSA \cite{EdDSA-RFC,Bernstein:2015}).
        \State Randomly generate a \( b \) bit string \( s \) 
        \State \( h \gets H(s) \) \Comment{Note that \( h = \left(h_0, \dotsc, h_{2b-1}\right)\)}
        \State \( a \gets 2^{b-1} + \sum_{i=c}^{b-2} h_i 2^i \)
        \State \( p \gets \sum_{i=0}^b h_{b+i}2^i\)
        \State \Return \( (a, p) \)
    \end{algorithmic}
\end{algorithm}


There are several possible approaches to parts of \cref{alg:distributed_eddsa_key_generation} which we discuss in the next subsections.
Each approach entails trade-offs between trust in the dealer and trust in the actor.
The choice of approach to take is consequently implementation (and use-case) dependent.

Note that all keys generated by \cref{alg:distributed_eddsa_key_generation} are keys that could be generated following either EdDSA or the analogous key sharing algorithms \cite{EdDSA-RFC}, and vice versa. It is important that the secret \( a_i \) be stored unreduced.
As illustrated in \cref{example:reduction}, after reduction modulo \( \ell \), the secret scalar potentially loses any small subgroup (or other) protection afforded by  clamping.

The simplest way to generate a secret scalar without a secret prefix is to randomly generate a value \( 2 \le a < \ell \) using a cryptographically secure generation method.
Alas, 
this removes the small subgroup protections afforded by key clamping, necessitating manual checks. Although, as discussed in \cref{ssub:preliminaries:keyclamping},  in the case of EdDSA, such protection may be unnecessary.

A similarly simple secret scalar generation that produces keys strictly in \( \mathit{KeySpace}_{25519} \) (or \( \mathit{KeySpace}_{448} \) for Ed448, or the analogous keyspace for any other curve) is:
\begin{enumerate}
    \item Generate \( 1 \le a < \lfloor{\nicefrac{\ell}{8}}\rfloor \)
    \item Calculate the secret as \( 8 a \) for Curve25519 (or \( 4 a \) for Curve448)
\end{enumerate}
Such keys would have small subgroup protection for key exchange however this method is inadvisable as it  significantly reduces the key space.

\begin{example}
\label{example:reduction}
 In the case of Ed25519,  observe that \( 2 \) can be realised as a value \( 8k \mod \ell_{25519} \) where \( 8k \) is a key clamped secret seed for Ed25519.
However, \( 2 \) cannot be generated as \( 8a \) for \( 1 \le a \le \lfloor{\nicefrac{\ell}{8}}\rfloor \).
\end{example}

The value \( 2 \) in \cref{example:reduction} is not unique; the same is true for \( a=4 \), \( a=5 \), \( a=7 \), \( a=10 \), and others.
Analogous examples for Ed448 (using \(4k\) and \(4a\) in place of \(8k\) and \(8a\) respectively) are easily found.

\subsection{Encryption Keys and Actor Identity}
\label{sub:threshold_keygen:encryption_keys_and_actor_identity}

Each actor, \( s_i \),  encrypts and decrypts messages to and from the other actors; see \cref{alg:distributed_eddsa_key_generation} \cref{alg:preliminary_key_generation:encrypt_share} and \cref{alg:distributed_eddsa_key_generation} \cref{alg:finalising_key_generation:decrypt_share}.

An  approach to actors knowing the encryption keys of their peers is for each actor, \( s_i \), to have a unique identifier in the system, \( \mathit{id}_i \) say.
Then actor $s_i$ has a public encryption key that is easily found with knowledge of \( \mathit{id}_i \), and potentially verifiable within the system (e.g., with the use of certificates).   A unique identifier allows knowledge of public keys for  peers to be outside the scope of our protocol; such identities, keys, and verification methods can be established at system creation.

An alternate approach is for the dealer to send public keys to the actors.
This would allow for a system in which actors do not have a system-wide identity, or systems for which such an identity is problematic. 
The dealer would need to send the public keys to the actors as part of \cref{alg:distributed_eddsa_key_generation:send_x_coords} of \cref{alg:distributed_eddsa_key_generation}.  If there are pre-existing keys for the actors, then the dealer could simply send them in addition to the \( x \)-coordinates in \cref{alg:distributed_eddsa_key_generation}, \cref{alg:distributed_eddsa_key_generation:send_x_coords}.

A system in which the actors have a pre-existing identity would imply that each actor, \( s_i \), knows (or could discover)  who their peers are when generating a key.
This, in turn, has implications for actor collusion, see \cref{sub:EdDSA:actor_collusion}.
There are also implications  for the choice of \( x \) coordinates for secret sharing, see \cref{sub:EdDSA:x_coordinates}.

It may be that ephemeral keys are desirable either in the absence of a system-wide identity or to prevent the actors from knowing the identities of their peers.
Incorporating ephemeral keys into \cref{alg:distributed_eddsa_key_generation} requires an extra round of communication: the notified actors  must generate the keys and send them to the dealer before the dealer sends the \( x \)-coordinates and keys to the actors (\cref{alg:distributed_eddsa_key_generation:send_x_coords} of \cref{alg:distributed_eddsa_key_generation}). 

We note that if the dealer sends encryption keys to the actors, then a malicious dealer could conduct a man in the middle attack.
If keys are to be sent by the dealer, then either the actors need to be able to verify the keys, or the dealer is necessarily absolutely trusted.

Underpinning the considerations discussed in this subsection is consideration of dealer trust and actor trust.
There is a complex interplay between these trust considerations and the logistical system decisions.
These considerations and ramifications are further explored in \cref{sub:EdDSA:actor_collusion}.

\subsection{Choice of Coordinates for Shamir Secret Sharing}
\label{sub:EdDSA:x_coordinates}

The \( x \)-coordinates (\(x_i\) from \cref{alg:distributed_eddsa_key_generation}) are related to, although separate from,  actor identity.
If actors have a static identity in the system, then those identities could be mapped to the \( x \)-coordinates. There could be a function \( \tilde{x} : \mathit{id}_i \mapsto x_i \) to produce an \( x \)-coordinate for secret sharing. 
For example, reduction modulo \( \ell \) is a simple option when \( \mathit{id}_s \) can be interpreted numerically.

An injective function \( \tilde{x} \)  ensures that no two actors, \( s_i \) and \( s_j \), have \( \tilde{x}(\mathit{id}_i) = \tilde{x}(\mathit{id}_j) \).
Conversely, if \( \tilde{x} \) is invertible, then the actors could infer the identities of their peers.

If deriving \( x_i \) deterministically from an identity is undesirable (or infeasible) then the dealer could  randomly generate $x$ coordinates and send them to the actors as part of the notification process.
This is perhaps the simplest method.

Note that the entropy of the \( x \)-coordinates is not important for the security of the Shamir secret sharing scheme. 
It is enough that the \( x \)-coordinates are unique (and in the correct ring).
Indeed, one could   use  \( 1, 2, \dotsc, n \).

In our system the actors generate the polynomial coefficients, and the entropy of the coefficients is protected by the nested scheme, as shown in \Cref{sec:RNG}.
So having the dealer generate the \( x_i \) is not a danger, even if the dealer has poor quality random number generation.
Having the actors, \( s_i \), randomly generate their own \( x_i \),  adds needless complications.

\subsection{Publication and Use of the Public Key}
\label{ssub:publicatoin_and_use_of_public_key}
The publication (or distribution) and use of the public key has profound effects on the systems resistance to the rogue key attack.
Ultimately there are two options: the dealer publishes/distributes the key, or the actors do (or both).
A related question is whether there is any validation of the key. 

We envision two classes of system: one in which the swarm (and the keys created) are part of a larger system, and one in which the swarm performs only generation, signing, and key exchange calculations at the behest of the dealer.

When part of a larger system,  integrity is of high importance.
The key needs to be verifiable, and the generation process auditable so that any rogue key attack attempts can be detected, and the keys not distributed.
Actors must independently publish the key to a log or ledger, and keys only be accepted by the system if all actors publish the same key.

When the swarm is merely a distributed (and more secure) replacement for locally created and controlled keys of the dealer, it will usually be the dealer that distributes the keys.
In other words, the swarm is an on-demand system for performing cryptographic calculations that would have otherwise been performed locally.

If the dealer publishes the key, then it is impossible to stop the dealer from conducting a form of  rogue key attack, despite the protections in our system.
We stress that the algorithm itself  presented here is \emph{not} susceptible to a rogue key attack (see \cref{ssub:discussion_keygen:rogue key attack prevention}).
In any system where the dealer publishes or otherwise distributes the public key, there is nothing to stop that dealer from making a new threshold key by combining a key they control in a direct threshold techniques with the aggregate public key from the swarm.
The dealer may, thus, perform a rogue key attack on that direct threshold key.

Note that such an attack produces nothing more than a dealer-local key (functionally identical to a locally generated key).

It is worth observing that that if the swarm is merely an on-demand extension of dealer-local keys, then the dealer might legitimately wish to combine a local key and the aggregate key in a direct-threshold key.
Such a key would prevent actors from colluding to reconstruct the secret scalar (see \cref{sub:EdDSA:actor_collusion}); any corrupt actors would require the dealer's local key to find the secret scalar corresponding to the published public key.

\section{Discussion}\label{sec:discussion}

\subsection{Swarm Random Number Generation Discussion} \label{subsec:swarmdiscussion}
Nested Shamir secret sharing is used as a multi-party random number generator,  dubbed Swarm RNG in \cref{sec:RNG}.  Multi-party RNG,  is not a new concept \cite{cleve1986limits}. The `coin tossing' protocols create streams of random bits \cite{cleve1986limits, beimel2015protocols} which can be used to create a random number.   The Swarm RNG protocol proposed in this paper generates a number within a range similar to \cite{syta2017scalable}.

The RandHerd protocol \cite{syta2017scalable} is a multi-party RNG based on Schnorr signatures.  RandHerd is not a coin flipping protocol, but rather can generate a random number form within a given set range like swarm RNG.  At the beginning of the RandHerd protocol a `dealer' is randomly elected. Provided that the dealer is honest, and the number of corrupted parties is below the threshold, the computation can compete and the random number is generated.  At the end of the RandHerd protocol, all parties learn the random number, whereas with Swarm RNG, the random number can remain secret.  The RandHerd protocol \cite{syta2017scalable} is designed for a public use case e.g. lottery numbers.   Swarm RNG is appropriate for a secret use case such as cryptographic key generation.

If more than half the parties are honest, then there exists multi-party coin tossing protocols with negligible bias suitable for cryptographic use.
When at least half the parties are corrupted, the bias in the multi-party coin tossing protocol is $O(\nicefrac{1}{r})$ for a protocol with $r$ rounds \cite{cleve1986limits}. Protocols have been developed to have low bias with assumptions on the number or proportion of corrupted parties in the system \cite{beimel2015protocols}.

\begin{theorem}\label{thm:unbiasedRNG}
Swarm RNG (as described in \cref{sec:RNG}) can achieve unbiased RNG as long as at least one of the parties uses an unbiased RNG to generate their share.
\end{theorem}
\begin{spiproof}
Let \(\varsigma\) be the secret random number which is generated as an aggregate of the secrets \(\varsigma_k\), for \(1\leq k\leq n\), according to the algorithm described in  \cref{sec:RNG}.  Each share \(\varsigma_k\) is generated independently by each party using their own RNG. 

Using \cref{eqn:entropy_multiaddition} the entropy of \(\varsigma\) is greater than or equal to the entropy of the highest entropy share.
\begin{equation}\label{eqn:entropy_secret}
  H(\varsigma) \ge \max \left\{H(\varsigma_1),H(\varsigma_2),\dots,H(\varsigma_n) \right\}
\end{equation}

Let the secret \(\varsigma_k\) be generated using an unbiased RNG for some \(1 \leq k\leq n\), then \(\varsigma_k\) has maximum entropy.   It follows from \cref{eqn:entropy_secret} that the aggregate secret, \(\varsigma=\sum_{i=1}^n \varsigma_i\), also has maximum entropy and so is unbiased.
\end{spiproof}




A party to the Swarm RNG may be faulty, or maliciously attempt a Denial of Service attack by withholding their share.

\begin{theorem}\label{thm:fair}
Swarm RNG (as described in \cref{sec:RNG}) is a fair multi-party computation.
\end{theorem}
\begin{proof}
Any party to the 
swarm RNG calculation may withhold their share, causing the dealer to abort the computation.  The dealer may choose a different set of parties and begin the key generation again.

Any party that withheld their share does not gain any information about the new swarm or new key generation. Thus, Swarm RNG is fair.
\end{proof}

With sufficiently many parties in the system, there are always enough to form a viable swarm and complete Swarm RNG.

Recently a protocol transformation was published which  uplifts a protocol from {\em fair} to {\em full security}  \cite{cohen2022fairness}.  
Starting with a fair protocol, and under the assumption that the fraction of honest parties is constant throughout the computation, the protocol can be transformed into a fully secure protocol.  Swarm RNG can be fully secure.


There are some other published distributed secret sharing schemes.
A dealer-less system removes the dealer as single point of vulnerability  \cite{kate2009distributed, das2021practical, Gennaro2007}, whereas \cref{alg:distributed_eddsa_key_generation} has a dealer without the dealer being a single point of vulnerability.
There are distributed RNG that have more steps than our system or require a broadcast channel \cite{cryptoeprint:2020/1390,Gennaro2007}.

We have not found any formal study of the entropy protection of using distributed RNG (Theorem \ref{thm:unbiasedRNG}) in prior literature.
A similar protection is mentioned for the uniform distributions of distributed RNG by Gennaro et. al. \cite{Gennaro2007}. 


\subsection{Threshold Key Generation Security Guarantees and Considerations}
\label{sec:discussion_keygen}

The threshold key generation scheme described in \cref{alg:distributed_eddsa_key_generation} (see \cref{sec:EdDSA}) fills the ``[TBS]'' for distributed key generation in the threshold modes IETF draft \cite[\textsection{4.4}]{draft-hallambaker-threshold-modes}. 
By generating a secret scalar using \cref{alg:distributed_eddsa_key_generation}, signing or key exchange may proceed according to the IETF drafts \cite{draft-hallambaker-threshold-modes,draft-hallambaker-threshold-sigs} (or, indeed, potentially any other threshold mode for EdDSA).

The aggregate secret scalar generated by \cref{alg:distributed_eddsa_key_generation} is \( a = \sum_{i=1}^n a_i \). Each actor \( s_i \) has a  secret share \( \sigma_i \) of $a$, but no party ever sees the aggregate secret scalar, $a$.



\subsubsection{Rogue Key Attack Prevention}
\label{ssub:discussion_keygen:rogue key attack prevention}
The zero-knowledge proof steps of \cref{alg:distributed_eddsa_key_generation} (creation on \cref{alg:preliminary_key_generation:create_zk_proof}, and verification on \cref{alg:finalising_key_generation:checks:low order curve point,alg:finalising_key_generation:checks:S,alg:finalising_key_generation:checks:verification equation}) guarantee that actor, \( s_i \), knows the secret scalar contribution, \( a_i \), corresponding to their public key contribution, \( \cvPt{A}_i \).
Moreover, this guarantee does not leak any information about the secret scalar contribution in question.

Inasmuch as our proof is an EdDSA signature of an empty message, we rely on the unforgeability properties of EdDSA \cite{brendel2021provable}.
We require  checks which are sufficient to guarantee both strong unforgeability (\cref{alg:distributed_eddsa_key_generation}~\cref{alg:finalising_key_generation:checks:S,alg:finalising_key_generation:checks:verification equation}, and message bound signatures (\cref{alg:distributed_eddsa_key_generation}~\cref{alg:finalising_key_generation:checks:low order curve point}).

A rogue key attacker (as discussed in \cref{ssub:Preliminary:ECC:Threshold Sigs and KEx:Attacks}) does not know the secret scalar corresponding to their public key contribution (without the ability to break the discrete logarithm problem).
As such, the guarantee that all actors can prove knowledge of their key contributions secret scalars eliminates the rogue key attack.
Even if a corrupt party sees all the public key contributions before calculating their own, they cannot both perform a rogue key attack, and provide a passing proof.

Note that the check for a low order curve point (\cref{alg:distributed_eddsa_key_generation}~\cref{alg:finalising_key_generation:checks:low order curve point}) prevents a rogue actor from providing a zero-knowledge proof that would otherwise pass if their \( \cvPt{A}_i \) was a low-order curve point. 
Such a forgery would require that \( H(\cvPt{A}_i,\cvPt{R}_i) \) is a multiple of \( 2^c \) (and so \( S_i = r_i \)), achievable with trial-and-error on the value of \( r_i\).

\subsubsection{Malicious Contribution}
\label{ssub:discussion_keygen:malicious contribution}

\Cref{alg:distributed_eddsa_key_generation}, as presented, produces a key for which we can prove that no rogue key attack has been performed (assuming the hardness of the discrete logarithm problem).
However there are two malicious contribution style attacks that it does not provide protections against.
Specifically, the actors, \( s_i \), cannot guarantee the legitimacy of the share contributions, \(\sigma_{j \to i}\), nor the public key contributions, \( \cvPt{A}_j\), sent by the other actors, \( s_j \).

For example, a malicious actor could generate and prove multiple \( \cvPt{A}_i \) contributions and send a different one---along with the corresponding proof of knowledge of \(a_i\)---to each other actor), or could send nonsense shares.

It is important to stress that such attacks do not compromise the security of the generated key.
Malicious contribution attacks will, however, cause the key generation to fail.
Such a failure might not be immediately apparent (as we similarly noted for key exchange \cref{ssub:Preliminary:ECC:Threshold Sigs and KEx:Attacks}---albeit for rogue key attacks).
If the actors publish to a log or ledger, as discussed in
\cref{ssub:publicatoin_and_use_of_public_key}, then malicious contribution attacks that lead to actors calculating different aggregate keys can be detected (although not the identity of the malicious actor(s)).

More rigorous checks such as Feldman secret sharing \cite{feldman1987} (as used in some dealerless key generation systems \cite{Gennaro2007,cryptoeprint:2020/1390}) require extra rounds and many more point multiplications.
The absence of a broadcast channel complicates the verifiable sharing as well; since all communication goes through the dealer, then a malicious dealer participating as one of the actors, or by colluding with a corrupt actor, can manipulate the public values used for the verification.
If actors are publishing the public key to a log or ledger (as we suggest for the aggregate key), then that could act as a broadcast channel for verifiable sharing.
The sharing schemes we've looked at that use verifiable sharing \cite{Gennaro2007,cryptoeprint:2020/1390} all have the property that malicious actors are ruled out, and the final group of shareholders may consist of between \( t \) and \( n \) members.

In \cref{alg:distributed_eddsa_key_generation} we have opted not to protect against a malicious contribution attack due to the low risk, compared to the extra communication and computation. 

A simple check would be to have each actor write \( \cvPt{A} \), and \( \sigma_i \cvPt{G} \) to a system log or ledger (independently of the dealer) and then a simple calculation checks the interpolated \( \cvPt{A} = \sum_{i=1}^n l_i \sigma_i \cvPt{G} \). The system may thus detect and reject keys for which the check does not hold.

Alternatively, if the key is to be used for signing, then a test signature could be produced instead.

\subsubsection{Actor Collusion}  
\label{sub:EdDSA:actor_collusion}

As observed in \cref{sub:preliminaries:threshold}, even though the actors do not need to reconstruct the secret in order to compute their signature or key exchange shares, a sufficiently large subset of actors may  collude to reconstruct the secret.
Even if the actors are not themselves malicious, an adversary who compromised enough actors may similarly reconstruct the key.
We note that such a compromise requires compromising several independent  systems, which is significantly harder than compromising a single target.
In addition, the adversary would need to know which actors to compromise in order to target a particular dealer.

We may wish to protect against the possibility of actor collusion.
We may observe that the actors (in addition to not needing to explicitly construct the secret key) do not even need to know their own $x$-coordinate, \( x_i \), in order to use the key. This observation can be stated formally:
\begin{remark}\label{prop:actor_xi}
Let $x_i$ be the   \( x \)-coordinate of actor $s_i$ as described in \cref{alg:distributed_eddsa_key_generation}. Then  actor $s_i$   does not  need knowledge of the value of \( x_i \) in order to sign messages \cite[\textsection 3.2]{draft-hallambaker-threshold-sigs} or exchange keys (see \cref{sub:preliminaries:threshold}).
\end{remark}

Note that \cref{prop:actor_xi} is not easily exploited to protect against actors colluding.
Signer, $s_i$ must know \( x_j \) in order to be able to calculate \( \sigma_{i \to j} \) (\cref{alg:distributed_eddsa_key_generation}, \cref{alg:preliminary_key_generation:calculate_share}).
Furthermore, actor $s_i$ must also know a key for actor \( s_j \) in order to encrypt the share.
Hiding the coordinate \( x_j \) from actors requires that each actor, \( s_i \), sends the entire polynomial \( P_i \) to the dealer, for the dealer to then calculate and encrypt the shares \( \sigma_{i \to j} \).
 This trivially allows the dealer to know the secret.

If the \( x \)-coordinate, \( x_j \), is separated from the identity of actor, \( s_j \), (see \cref{sub:EdDSA:x_coordinates}) then it may seem that the actors do not know which \( x \)-coordinate is their own.
However, the need to encrypt the shares allows actor \( s_i \) to identify their own \( x \)-coordinate, as they could recognise their own encryption key.

We note that in \cref{alg:distributed_eddsa_key_generation} the actors do not use the \( x \)-coordinates outside of preliminary key generation (the \Call{Complete}{$\cvPt{A}, \sigma_{1 \to i}, \dotsc, \sigma_{n \to i}$} procedure in \cref{alg:distributed_eddsa_key_generation}).
If all actors destroy these coordinates after use, then  the secret cannot be reconstructed by any colluding actors.
However, those \( x \)-coordinates are needed by the dealer in order to construct the Lagrange coefficients \( l_i\), so they must either be stored by the dealer or be deterministic. 
Storage of the coordinates introduces extra complications for the dealer, similar to those introduced by having a dealer-local signing key.
Deterministic coordinates implies that actors can infer their own \( x \)-coordinate.

Even if an actor is honest at the time of key generation and does not store any of the \( x \)-coordinates (including, in particular, their own \( x \)-coordinate), then that actor might still be able to later collude with other actors to reconstruct the secret.
If even one corrupt actor saves the entire list of \( x \)-coordinates at key generation time, then there are \( \binom{n-1}{t-1} \) combinations to try (assuming that the dishonest actor additionally noted which \( x \) coordinate was their own).
Each additional dishonest actor reduces this search space.
If we have \( d \) dishonest actors at key generation time, then there will be \( \binom{n-d}{t-d} \) combinations to search. 
If \( d \ge t \) then the secret can be directly reconstructed.

\begin{proposition} \label{prop:collusion}
Suppose that \cref{alg:distributed_eddsa_key_generation} is implemented such that the actors do not store their own \(x\)-coordinates. Let $d$ be the number of corrupt actors during the key generation phase of the scheme and suppose that those actors retain knowledge of the \( x \)-coordinates.
Further suppose that \( d < t \).
    
If \( t-d \) actors are later corrupted, then the resulting \( t \) corrupted actors can together reconstruct the key by searching \( \binom{n-d}{t-d} \) combinations of \( x \)-coordinates.

\end{proposition}
\begin{proof}
  Assume that all corrupt actors are colluding.
  Denote the \( d \) corrupt actors from key generation as \( s_{1}, \dots, s_{d} \).
 Denote the newly corrupted \( (t-d) \) actors as \( s_{d+1}, \dotsc, s_{t} \). 
 
    Actors \( s_{1}, \dots, s_{d} \) know the entire set of \( n \) \( x \)-coordinates and which coordinates correspond to \( x_{1}, \dotsc, x_{d} \).
    None of the actors know which of the \( x \) coordinates correspond to actors  \( s_{d+1}, \dotsc, s_{t} \).
    There are \( \binom{n-d}{t-d} \) ways to assign the unassigned \( t-d \) coordinates to actors \( s_{d+1}, \dotsc, s_t \).
    The actors can try all combinations.
\end{proof}
If the swarm size, $n$, is kept reasonably small (e.g., $n=20$), then the search space needed for the attack in \cref{prop:collusion} is feasible.
A set of honest actors at key generation prevents this attack.

\begin{corollary}
Consider \cref{alg:distributed_eddsa_key_generation}. If all parties are honest at the time of key generation, then reconstruction of the key is not possible.
\end{corollary}





Ultimately there is a trade-off between complications and the risk of dishonest actors being able to reconstruct the key:
\begin{itemize}
\item Greatest protection against the aggregate key being reconstructed by any party is achieved by having a dealer-local signing key but introduces extra complexity for the dealer.
\item Greatest protection against actors reconstructing the secret can be achieved with random \( x \)-coordinates and ephemeral signing keys, but comes at the expense of extra communication round trips, and guarantees that a dishonest dealer can discover the aggregate key.
\item Greatest protection against the dealer being able to reconstruct the key is achieved by having pre-existing identities and keys for the actors but allows sufficiently many malicious actors (either intrinsically malicious, or later compromised) to be able to reconstruct the key.
\end{itemize}

\section{Conclusion and future directions \label{sec:conc}}

The use of nested Shamir secret sharing for threshold elliptic curve encryption  provides high-quality cryptographically secure secret keys without any single point of vulnerability. Furthermore, this contribution addresses the missing distributed key generation specification in the threshold modes IETF draft specifications \cite{draft-hallambaker-threshold-modes}.

We have focused on twisted Edward curve based algorithms; the algorithm is possibly adaptable to other elliptic curve technologies (e.g., ECDSA) or even non elliptic curve schemes such as RSA. We do not comment further on this as we have not considered the necessary tight coupling of nested Shamir secret sharing for these algorithms. 

The Tide foundation developed the mechanism of manifesting unobtainable secrets using nested Shamir secret sharing as a key generation method in the context of a broader zero-trust, decentralised authority system \cite{TIDE}. The system is designed as new capability for digital platforms to manage verifiable authority over accounts, data and other resources in the form of programmable keys that no single entity ever has custody over \cite{hallvermaskerritt2022}.


\subsection{Future directions}
\subsubsection{Extension to Post Quantum Cryptographic Algorithms}
Quantum computers are currently laboratory experimental set ups. However, there is the potential for a commercial quantum computers  in the next 10 years.  We need to consider migration to quantum-safe cryptographic algorithms \cite{NIST2022status}, as algorithms have already been developed that significantly speed up the calculation of a discrete logarithm,  the heart of Elliptic curve cryptography.

In July 2022,  NIST announced three candidates for  quantum safe signature algorithms and one Key Encapsulation Mechanism for standardisation \cite{NIST2022status}, and a further four Key Encapsulation Mechanisms for further study.  They use the mathematical primitives of lattices, hashes, error correcting codes and elliptic curve isogonies. A fruitful area of investigation is to examine these `quantum safe' ingredients for  use with swarm RNG and threshold signing.

Although a flaw was found in the isogony based Key Encapsulation Mechanism algorithm \cite{castryck2022efficient}, it may still be possible to use isogonies in a threshold signature scheme.  It is a difficult problem, \cite{cozzo2019sharing}  still under research.





\subsubsection{Extending the IETF drafts}
The threshold signatures IETF draft recommends using randomly generated secret signing nonces (as we discussed in \cref{sub:preliminaries:threshold}). This is a deviation from the regular specification.
Using a deterministic method of calculating the secret signing nonce, allows the dealer to discover key shares by requesting multiple signature contributions from each signing party \cite[\textsection{}3.4.1]{draft-hallambaker-threshold-sigs} (and discussed in \cref{sub:preliminaries:threshold}).  Deterministic nonces are used in the EdDSA standard \cite{EdDSA-RFC}.  Future work includes exploring deterministic nonce generating methods for use in threshold signatures that are not vulnerable to dealer manipulation, or other known attacks.


\section*{Acknowledgements}
The Tide foundation is a company commercialising the concepts presented in this paper.
We acknowledge that the Tide foundation developed nested Shamir secret sharing.
The authors YH, ML, and DV work for the Tide foundation and collaborated with authors JH, MS, and GV from RMIT University to validate and develop the ideas which have been presented in this paper.

We acknowledge the contribution of Jose Luis Lobo in extensive technical discussions at the beginning of this project.

\bibliography{TIDEBibliography}


\end{document}